\documentclass[letterpaper]{aa}
\usepackage{natbib}
\bibpunct{(}{)}{;}{a}{}{,}
\usepackage{graphicx}

\begin{document}
\title{Are the sungrazing comets the inner source of pickup ions and energetic neutral atoms?}
\author{M. Bzowski and M. Kr{\'o}likowska}
\offprints{M. Bzowski, \email{bzowski@cbk.waw.pl}}
\institute{Space Research Centre PAS, Bartycka 18A, Warsaw, Poland}
\date{Received 28 April 2004 / Accepted 31 January 2005}

\newcommand{\asr}{Adv. Space Res.}
\newcommand{\ag}{Ann. Geophys.}
\newcommand{\pss}{Planet. Space Sci.}
\newcommand{\aasupp}{Astron. Astrophys. Suppl. Ser.}
\abstract{ We present arguments that at least part of the inner
source of pickup ions in the solar wind might be the material
released by sungrazing comets. Based on a statistical analysis of
sungrazing comets detected over almost eight years of LASCO
operation (1996 -- September 09 2004) an overwhelming majority of
the observed sungrazers belong to the Kreutz group of comets,
follows tightly clumped orbits and break up at $\sim 40 - 4$ solar
radii in a well defined region of space. The material released
from these comets could be (after ionization) an important portion
of the inner source of pickup ions (PUIs), as the local mass flux
of the inner source and cometary PUIs seem comparable. We indicate
time intervals during the year when the cometary PUIs could be
observed from a spacecraft on the Earth's orbit (from the end of
July until the end of the year) and show three time intervals when
they should be observable by Ulysses (from its launch time until
the end of 1990, from the end of November 1994 until mid-May 1995
and from February 2001 until the end of July, 2001). We argue that
the PUIs from the inner source should include both singly and
doubly charged ions and that this cometary hypothesis alleviates
some difficulties (in particular, the issue of hydrogen deficit)
in the interpretation of the inner source as solar wind
neutralized on dust grains close to the Sun. }

\titlerunning{Sungrazing comets as inner source of pickup ions}

\authorrunning{M. Bzowski \& M. Kr{\'o}likowska}
\maketitle

\section{Introduction}
Pickup ions (PUI) in the solar wind are former neutral atoms of
thermal energy, ionized and picked up by the solar wind (Fahr
1973\nocite{fahr:73}). He$^+$ PUI created from interstellar atoms
penetrating the heliosphere were first detected by M{\"o}bius et
al. (1985)\nocite{mobius_etal:85a} and H$^+$ by Gloeckler et al.
(1993)\nocite{gloeckler_etal:93a}. The inner source of pickup ions
was discovered by Geiss et al. (1994, 1995).
\nocite{geiss_etal:94a, geiss_etal:95a} Its elemental composition,
which includes oxygen (mass 16), nitrogen (mass 14), neon (mass
20), hydrogen, carbon (mass 12), magnesium (mass 24) and silicon
(mass 28), identified by the mass/charge ratio (Gloeckler \&
Geiss, 1998\nocite{gloeckler_geiss:98a}; Gloeckler et al.,
2000a\nocite{gloeckler_etal:00a}), is quite similar to the solar
wind composition. This suggests that the inner source is related
to dust or that this portion of the PUI population must be solar
wind ions neutralized, slowed down and reionized. Analysis of the
inner source PUI velocity distribution function indicates that the
source must be located very close to the Sun ($< 0.1$~AU).

In this communication we suggest that at least part of the inner source
of pickup ions might be the material released by sungrazing comets.

Shortly after beginning of operations, the LASCO coronograph
on-board the SOHO spacecraft discovered a stream of comets
approaching the Sun to a few solar radii (Biesecker et al.
2002\nocite{biesecker_etal:02}). Until early September 2004,
almost 850 such objects have been observed and orbits of 809 have
been determined.

We present statistical analysis of these 809 SOHO sungrazers based
on the orbital data taken from the Catalogue of Cometary Orbits
(Marsden \& Williams, 2003\nocite{marsden_williams:03}) and from
IAU Circulars publicly available on the Web at
http://cfa-www.harvard.edu/iauc/RecentIAUCS.html (Cambridge, USA).
We confirm the earlier finding by Biesecker et al.
(2002)\nocite{biesecker_etal:02} that the sungrazers are in fact
more frequent than observed and we postulate that due to this high
frequency of apparitions they should provide a source of pickup
ions which is virtually ever-present and well-constrained in
space. We indicate time intervals during the year when the
sungrazer comet-related PUI should be observable by Earth- (and
L1)- bound spacecraft and the portions of the Ulysses orbit where
these ions should reach the spacecraft. We discuss Energetic
Neutral Atoms (ENA) created due to charge exchange between solar
wind protons and cometary neutral atoms and we conclude that owing
to reionization in the proximity of the Sun they should create an
extra source of ``inner source'' pickup protons.

\begin{figure*}
\includegraphics[width=17cm]{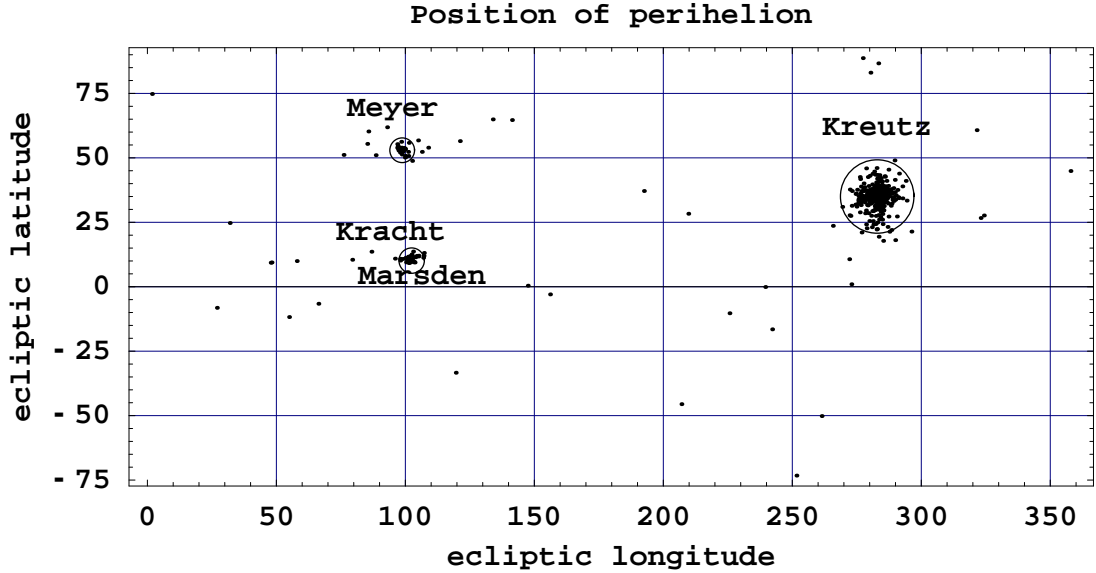}
\caption{Distribution of ecliptic coordinates of perihelion points
of known SOHO sungrazers. For October 2004, the primary Kreutz
group has 686 members, the Meyer group 40 and the Marsden \&
Kracht group together 39. Circles mark the angular distances
around respective mean values containing 99\% of the members of
the three groups.} \label{fig1}
\end{figure*}

\begin{figure*}
\centering
\includegraphics[width=17cm]{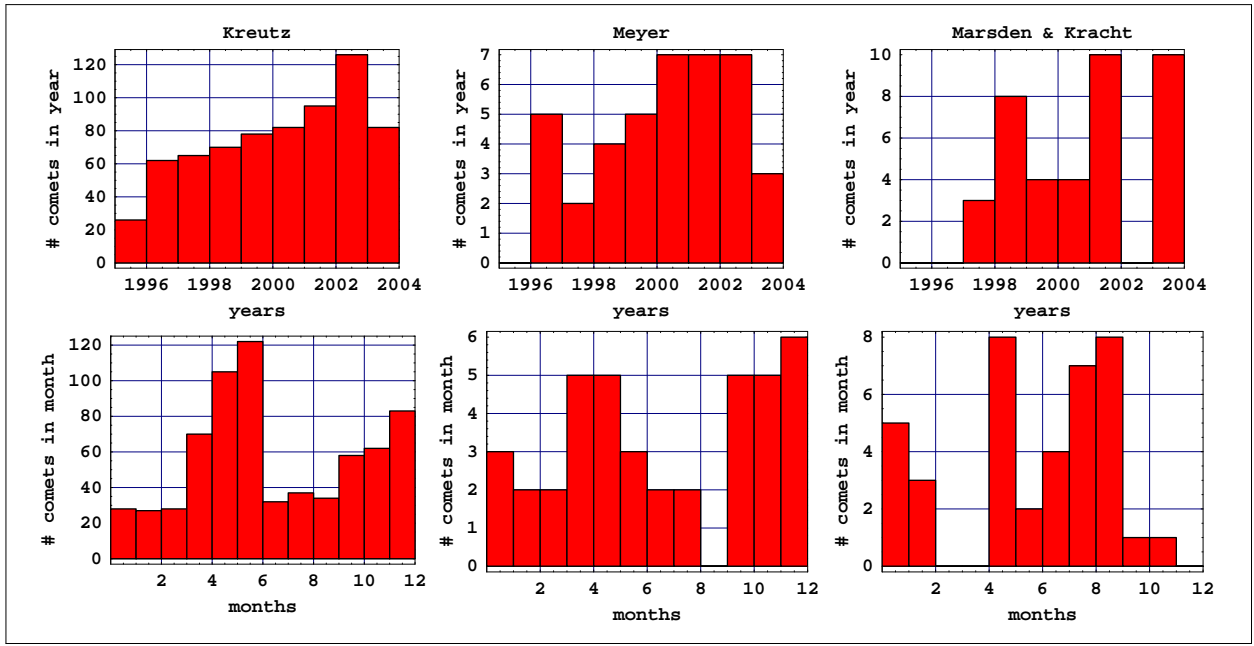}
\caption{Histograms of perihelion passage times for the SOHO
sungrazing groups of comets: upper panel -- yearly, lower panel --
monthly. The pattern for the monthly rates of the Kreutz group holds
for each year within statistical uncertainty.}
\label{fig2}
\end{figure*}

\section{Statistics of orbits and perihelion time of sungrazing comets}

An overwhelming portion (86$\%$) of the sungrazing comets belongs
to the Kreutz group. The Kreutz group of comets was discovered in
the 19-th century (Kreutz 1888, 1891,
1901\nocite{kreutz:1888}\nocite{kreutz:1891}\nocite{kreutz:1901})
and observed also later throughout the 20-th century: visually
(Marsden 1967\nocite{marsden:67}) and by space-borne SOLWIND, SMM
(Sheeley et al. 1982, \nocite{sheeley_etal:82} Michels et al.
1982,\nocite{michels_etal:82} Marsden 1989\nocite{marsden:89},
MacQueen \& StCyr 1991\nocite{macqueen_stcyr:91}), and LASCO
coronographs. The pre-LASCO observations discovered 46 sungrazing
comets. After the initial period of LASCO observations, the
discovery rate quickly increased and since 1997 it has been
oscillating between 60 and 120 comets per year with orbital
elements determined (Fig. \ref{fig2}, upper-left panel; see also
Biesecker et al, 2002\nocite{biesecker_etal:02}). Hence we believe
that the Kreutz group has been a persistent phenomenon at least
during the era of in situ space exploration, presumably with a
weakly changing annual rate.

In addition to the Kreutz group, there are three additional groups
of sungrazing comets, by far less numerous than the Kreutz group
but nevertheless non-negligible as a potential source of the
``inner source'' PUIs: Meyer (5$\%$ of all SOHO sungrazers),
Marsden (2$\%$), and Kracht (3$\%$). Their orbital parameters seem
correlated in a certain way with the Kreutz group. These four
groups include almost all known sungrazing comets (except 35
``stragglers'').

\begin{table}
\caption{Mean and most probable orbital parameters of the Kreutz
group of sungrazing comets; total of 686 comets in the sample.}
\label{tab1}
\begin{tabular}{lcccc}
  \hline
  Parameter                         & Value      & Std.       & Most     & Mean from \\
                                    &            & Dev.       & probable & Sekanina \\
  \hline
  Perihelion longitude [\degr]      & 282.9      & 2.9                 & 283. &  282.6\\
  Perihelion latitude  [\degr]      & 35.0       & 3.4                 & 35.  &  35.2 \\
  Perihelion distance  [$R_{\sun}$] & 1.22       & 0.22                & 1.07 &  1.10\\
  Inclination          [\degr]      & 143.5      & 2.8                 & 144. &  144.5 \\
  Perihelion argument  [\degr]      & 80.1       & 12.0                & 85.  &  83. \\
  Node                 [\degr]      & 1.0        & 14.4                & 7.  &  4. \\
  \hline
\end{tabular}
\end{table}

\begin{table}
\caption{Mean orbital parameters of the Meyer and Marsden \&
Kracht groups of sungrazing comets (respectively, a total of 40
and 39 comets in the samples).} \label{tab2}
\begin{tabular}{lcc}
  \hline
  \multicolumn{3}{c}{{\it Meyer Group}}                  \\
  Parameter                         & Value      & Std. Dev. \\
  \hline
  Perihelion longitude [\degr]      & 98.8       & 1.3   \\
  Perihelion latitude  [\degr]      & 53.0       & 1.4   \\
  Perihelion distance  [$R_{\odot}$]& 7.7        & 0.4  \\
  Inclination          [\degr]      & 72.2       & 1.3   \\
  Perihelion argument  [\degr]      & 57.0       & 1.2   \\
  Node                 [\degr]      & 73.6       & 1.3   \\
  \hline \hline
  \multicolumn{3}{c}{{\it Marsden \& Kracht Group}}      \\
  Parameter                         & Value      & Std. Dev. \\
  \hline
  Perihelion longitude [\degr]      & 101.9      & 2.4   \\
  Perihelion latitude  [\degr]      & 10.8       & 1.1   \\
  Perihelion distance  [$R_{\odot}$]& 9.90       & 0.96  \\
  Inclination          [\degr]      & 18.5       & 6.9   \\
  Perihelion argument  [\degr]      & 45.5       & 19.0   \\
  Node                 [\degr]      & 57.7       & 20.0   \\
  \hline

\end{tabular}
\end{table}

Comets from the Kreutz group follow extremely elongated elliptical
orbits (Sekanina 2002a\nocite{sekanina:02a}). Because of poor
quality of positional data, all orbits of SOHO sungrazers could
only be determined with the assumption that their eccentricity is
equal to 1. The ecliptic coordinates of perihelia of all LASCO
sungrazers with known orbits, detected before September~9~2004,
are shown in Fig. \ref{fig1} and their mean orbital elements in
Tables \ref{tab1} and~\ref{tab2}. The Marsden and Kracht groups,
although slightly different to each other, have such similar
trajectories close to the Sun that in this paper we will treat
them together as the Marsden \& Kracht group. Since there were no
reports on apparitions of comets from these groups before LASCO,
it may be -- especially in the case of the Marsden \& Kracht group
-- that they are transient phenomenon lasting just a few years. On
the other hand, even the Kreutz comets were observed rarely and
far in between before LASCO, so it cannot be ruled out that the
lack of earlier apparitions is a selection effect.

The yearly rate of sungrazing comets from all groups is most
probably biased by selection effects related to the construction
of the LASCO coronographs (Biesecker et al.
2002\nocite{biesecker_etal:02}). When binned monthly, all groups
show a statistically significant seasonal variability of the
apparition rate, featuring one sharp peak and a broader secondary
peak 6 months later (Fig. \ref{fig2}, lower panel). For the Kreutz
group, the monthly pattern shown in Fig. \ref{fig2} persists for
each separate year (within deviations due to the duty cycle of the
LASCO instrument and statistical spread); the other groups are too
scarce for a statistically significant analysis of this phenomenon
on a yearly basis. The primary peak of the Kreutz group occurs in
June and the secondary in December (see Fig. \ref{fig2},
lower-left panel). The monthly peaks of the Meyer group correspond
well with the Kreutz peaks (Fig. \ref{fig2}, second panel in the
lower row), while the peaks of the Marsden \& Kracht group seem to
be shifted in time by about 2 months (Fig. \ref{fig2}, lower right
panel).

The highest monthly rate during the whole LASCO observation period
occurs in June and we adopt this rate as the true rate of the
Kreutz group. During the seven full years of LASCO observations
(1997 -- 2003) about 100 comets were observed during this month.
Thus we conclude that the true apparition rate is $100:7 \simeq
14$ per month, or one every other day. Based on a much smaller
sample, observed from the end of 1996 until the end of 1998,
Sekanina (2003\nocite{sekanina:03}) gives a similar estimate for
the apparition rate: 0.6 per day. The peak monthly rates for the
Meyer and Marsden \& Kracht groups are, correspondingly, 5 and 8
during the $\sim 7$ years period, which yields about one per month
in each of the two groups.

Based on the lightcurves of $\sim 20$ brightest objects, Sekanina
(2003\nocite{sekanina:03}) gives an estimate of the number $N_{\rm cum}$ of
comets with the mass greater than or equal to $M$ as:
\begin{equation}
\label{eq0}
N_{\rm cum}\left(M\right) = \left(\frac{M_0}{M}\right)^\nu = 22 \left(\frac{10^{10}\,{\rm
g}}{M}\right)^{0.68}
\end{equation}
(with the selection effects taken into account) and of the total
mass $M_{\rm sum}$ carried by the train of Kreutz comets with
masses lower than or equal to $M$ as:
\begin{equation}
\label{eq1}
M_{\rm sum}\left(M\right) =
\frac{\nu}{1-\nu}\,M_0\,\left(\frac{M_0}{M}\right)^{\nu-1} .
\end{equation}
In these equations, $M_0$ is an unknown  mass of the most massive
comet observed and $\nu = 0.68$ is a population index fitted by
Sekanina based on the lightcurves observed. We find the mass of
the most massive comet $M_0$ from Eq. (\ref{eq0}) by substituting
$N_{\rm cum} = 1$ and we obtain $M_0 = 0.94\cdot 10^{12}$~g. The
total mass of the incoming comets is thus computed as $M_{\rm
sum}\left( M_0 \right)$ and given by the formula:
\begin{equation}
\label{eq1a}
M_{\rm sum} = M_0\, \nu / \left(1 - \nu \right).
\end{equation}
This gives  $M_{\rm sum} = 2\cdot 10^{12}$~g
arrived during the $\sim 2$ years long interval, equivalent to
$3.1\cdot 10^4$~g~s$^{-1}$.

Modelling by Iseli et al. (2002\nocite{iseli_etal:02}) returns the
most massive element at $0.63 \cdot 10^{12}$~g, which agrees with
the observational value to $\sim30$\%. However, observations of
several bright sungrazers by Raymond et al.
(1998)\nocite{raymond_etal:98} and Uzzo et al.
(2001)\nocite{uzzo_etal:01} suggest that their diameters are
5-fold smaller than inferred by Sekanina and Iseli, which
indicates a substantially smaller mass. Since the width of the
Ly-$\alpha$ line observed by UVCS suggests that those actually
seen are not the original H-atoms from the comet but solar wind
protons charge-exchanged with cometary atoms, we adopt results of
the direct estimates, $3.1\cdot 10^4$~g~s$^{-1}$, by Sekanina
(2003\nocite{sekanina:03}).

Typical sungrazing comets disintegrate between 40 and $4\,
R_{\sun}$ (0.19 -- 0.019~AU), typically at $10\, R_{\sun} \simeq
0.05$~AU (Sekanina 2003\nocite{sekanina:03}). Very few of the
comets (if any) make it through perihelion. After breakup, almost
the whole mass is transferred to the solar wind as pickup ions.
Additionally, the charge exchange ionization with solar wind
protons produces Energetic Neutral H Atoms (H ENA) at energies
from $\sim 30$~eV to $\sim 0.8$~keV.

\section{Opportunities for pickup ion and ENA observations}
\subsection{Pickup ions}

\begin{figure*}
\centering
\includegraphics[width=17cm]{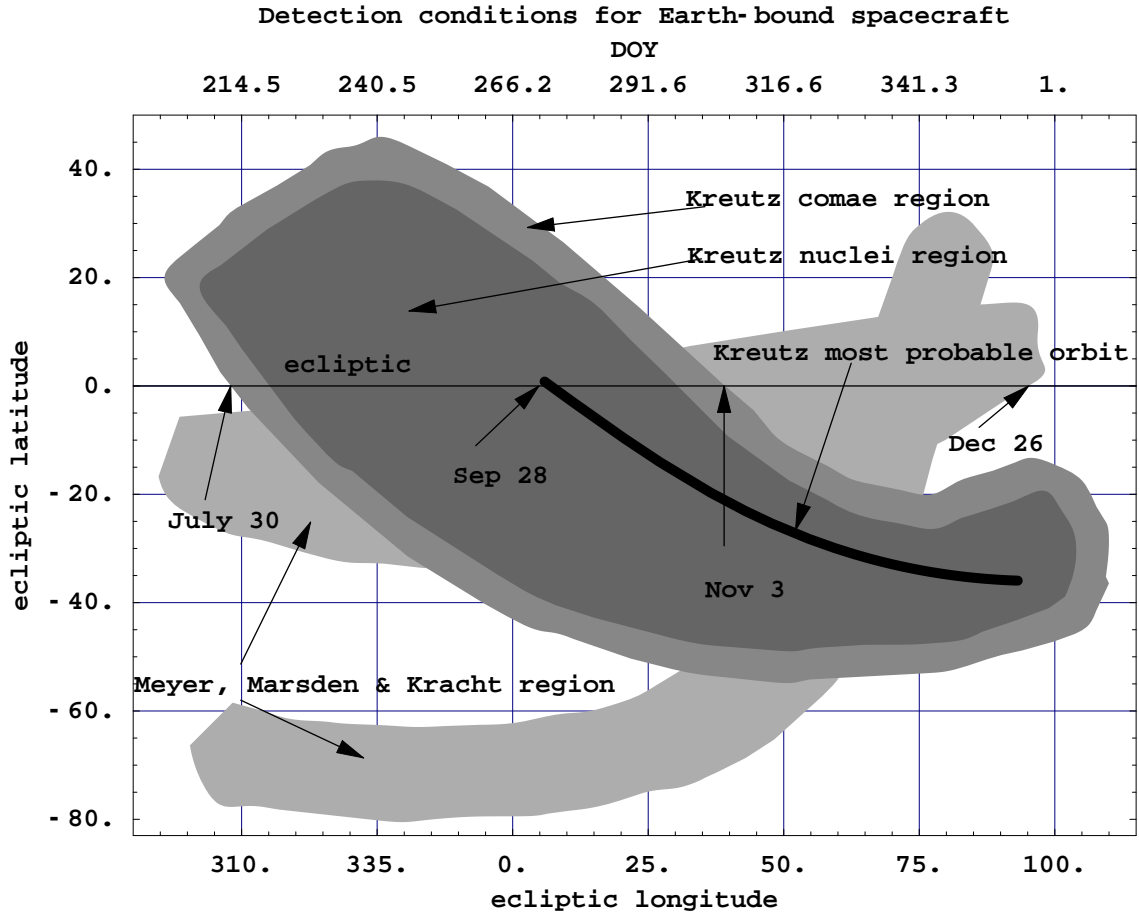}
\caption{Detection chart for sungrazing comet-related PUI and ENA
by Earth-bound spacecraft. The dark region is a projection on sky
of the volume penetrated by nuclei of the Kreutz group comets. Its
lighter envelope is a projection of the volume filled by their
comae. The light gray streaks are sky projections of volumes
penetrated by Meyer and Marsden \& Kracht comets (with the comae).
The black line is a sky projection of the most probable orbit of
the Kreutz group. Arrows mark the dates of Earth entering and
exiting the region prone for detection of the sungrazing comets
populations and the date of the intersection with the projection
of the most probable Kreutz orbit. The lower horizontal axis is
scaled in ecliptic longitude and the upper axis in the
corresponding DOY values; the verical axis is ecliptic latitude.}
\label{fig3}
\end{figure*}

\begin{figure*}
\centering
\sidecaption
\includegraphics[width=17cm]{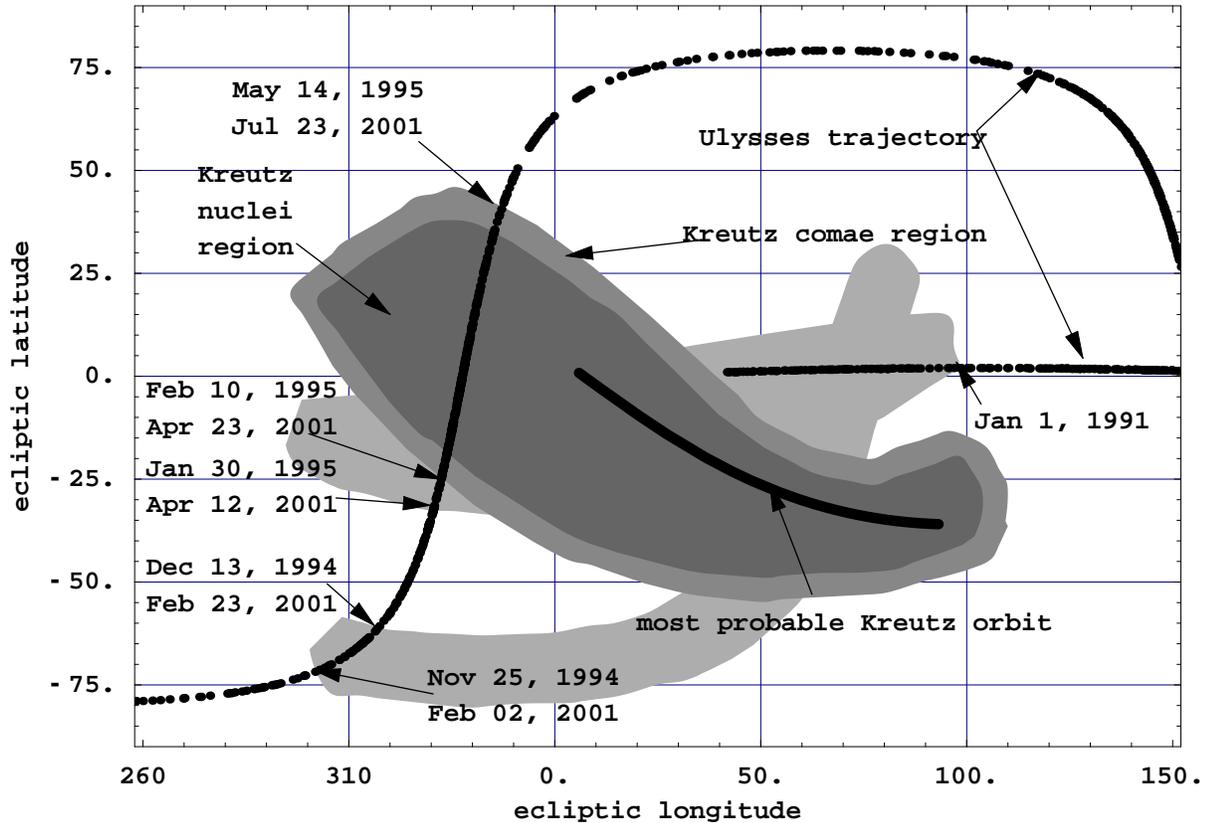}
\caption{Detection chart for the sungrazing comet-related PUI by
Ulysses. The gray regions are identical as in Fig. \ref{fig3}. The
Ulysses trajectory is marked with the dotted line. Ulysses was
launched while in the Meyer, Marsden \& Kracht region and exited
it on January 1, 1991. Then, on its two out-of-ecliptic
revolutions around the Sun, it crossed the regions prone  to
detection of the sungrazer population on the days indicated in the
figure.  } \label{fig4}
\end{figure*}

\begin{table*}
\caption{Detection intervals for sungrazing comet-related PUI and
ENA for Earth-bound spacecraft.$^{\rm a}$} \label{tab3}
\begin{tabular}{lcccc}
\hline
  Comet group                 &  Kreutz     & Kreutz               & Kreutz       &   Meyer, Marsden \& Kracht \\
  event                       & in          &  maximum             & out          &   out        \\
\hline
  Date                        & July 30     &  Sep 28              & Nov 3
& Dec 26  \\   DOY                         & 211         &  271 & 307          &
360   \\   Ecliptic longitude  [\degr] & 306         &  5 & 39           & 95
\\ \hline  \small{$^{\rm a}$ see also Fig.~3} &&&& \\
\end{tabular}
\end{table*}

\begin{table*}
\caption{Detection intervals for sungrazing comet-related PUI and
ENA for Ulysses.$^{\rm a}$} \label{tab4}
\begin{tabular}{lccccc}
\hline
  Comet group                 &   Meyer    & Meyer          & Marsden \& Kracht &  Kreutz       &   Kreutz      \\
  event                       &   in       &  out           &  in               &   in          &  out          \\
\hline
  Date                        &  Nov 25    &  Dec 13, 1994  &  Jan 30, 1995     & Feb 10, 1995  & May 14, 1995  \\
  DOY                         &  329       &  347           &  30               & 41            & 134           \\
  Ecliptic latitude [\degr]   &  $-$72~~   &  $-$60~~       &  $-$33~~          & $-$26~~       &  41           \\
&&&&& \\
  Date                        &   Feb 02   &  23, 2001      &  Apr 12, 2001     &  Apr 23, 2001 & July 23, 2001 \\
  DOY                         &   33       &  54            &  102              & 113           & 204           \\
  Ecliptic latitude [\degr]   &   $-$72~~  &  $-$60~~       &  $-$33~~          & $-$26~~       &  41           \\
   \hline
  \small{$^{\rm a}$ see also Fig.~4} &&&&& \\
\end{tabular}
\end{table*}

\subsubsection{Seed population}

The material released from comets during the approach phase
includes water molecules and CO$_2$ and CO at the abundance of 3
-- 10\% and 0.5 -- 20\% with respect to water. Other molecules and
atoms are much less abundant, below 1\% by number (Geiss \& Altweg
1998\nocite{geiss_altweg:98}, Bocl{\'e}e-Morvan et al.
2004\nocite{boclee-morvan_etal:04}). Hence the cometary PUIs from
the approach phase should be almost entirely hydrogen and oxygen
ions and the number abundance of oxygen with respect to hydrogen
prior to ionization  should be roughly 0.5.

The mass composition of cometary nuclei given by Greenberg \& Li
(1999\nocite{greenberg_li:99}) involves $\sim 26$\% of water,
$\sim 23$\% of organic refractory material (dominated by carbon),
$\sim 9$\% of carbon itself and $\sim 26$\% of silicates, whose
chemical composition can be approximated by [Mg,Fe]$_2$SiO$_4$.
CO, CO$_2$, CH$_3$OH and H$_2$CO make up $\sim 11$\% the of
nucleus by mass, and all other add up to $\sim 5$\%. Based on
these data we estimate that the nucleus is composed of $\sim 43$\%
H atoms, $\sim 27$\%  O atoms and $\sim 25$\% C atoms; magnesium,
silicon, iron and other atoms are only $\sim 5$\% of the total
number density.

The dissociation rates of the cometary molecules are usually much
higher than the ionization rates of ions, so one expects that all
the material injected into the solar wind in the breakup region
(including also the dust grains, Mann et al.
2004\nocite{mann_etal:04}) will be totally dissolved into atomic
ions. Considering this, and the atomic weights, one obtains the
atomic production rate equal to $\sim 2\cdot 10^{27}$~s$^{-1}$.
Iseli et al. (2002\nocite{iseli_etal:02}) estimate that oxygen and
carbon from the breakup region when observed at 1~AU should be in
high ionization states because of EUV secondary ionization after
pickup. On the other hand, in the case of material injected into
the solar wind during the approach phase, i.e., at larger
heliocentric distances, one can expect to observe some molecules
(mainly water and hydrocarbon dissociation products) and the
probability of survival of PUIs in the singly charged state is
higher.

The composition of the material released during the breakup should
differ from the composition of volatiles released during the
approach phase. The cometary material also involves dust; in fact
most of the heavy elements in cometary nuclei are locked in dust
grains, which are released from the nucleus into the surrounding
space but continue towards the Sun. Some researchers hypothesize
that sungrazing comets may be the main source of dust very close
to the Sun, inside $\sim 3 R_{\sun}$ (Mann et al.
2004\nocite{mann_etal:04}). Those dust grains will interact with
the solar wind and solar radiation and additional atoms and
molecules will be inserted into the solar wind. ``Fresh'' cometary
dust, especially released from inside the breaking nucleus at a
few solar radii should sublimate quickly to become
dissociated/ionized and picked up by the solar wind. Kimura et al.
(2002)\nocite{kimura_etal:02} suggest that the dust observed in
the tails of sungrazing comets consists mainly of crystalline
silicates and perhaps also of amorphous pyroxene at distances
farther than $20\,R_{\sun}$. This is consistent with the
suggestion by Greenberg \& Li (1999)\nocite{greenberg_li:99} that
cometary nuclei contain dust grains of a silicate core and organic
refractory mantle with external mantles of water ice with
carbonaceous and PAH particles embedded; very close to the Sun the
mantles sublimate almost immediately after release of the grain
from the nucleus, and the ``rocky'' silicate cores sublimate
between 4 and $1.5\, R_{\sun}$ (Mann et al.
2004\nocite{mann_etal:04}). This is about the distance of
perihelia of the sungrazing comets. Hence we conclude that even
the most resistant dust cores will not survive the perihelion
passage and the entire cometary material will be evacuated with
the solar wind as pickup ions.

\subsubsection{Ionization and pickup}

The molecules released from the nucleus with a relative velocity
of $\sim 1$~km~s$^{-1}$ continue on the original orbits until they
become dissociated by solar radiation and then ionized by
photoionization, electron impact or charge exchange with the solar
wind. Once created, the new ions from the cometary material are
immediately picked up by solar wind and transported away from the
Sun.

Because orbits of most of the sungrazing comets are tightly
clumped in space, the region where the PUIs are created is well
defined. Comets are known to release gas and dust at $\sim 5$~AU,
the aphelion of the Ulysses orbit. Thus the source region for
sungrazer PUIs will be the volume of space traversed by the Kreutz
comets, extended by the typical size of cometary neutral gas
clouds, which is of the order of the Sun's diameter (M\"{a}kinen
et al. 2001a,b,\nocite{makinen_etal:01a} \nocite{makinen_etal:01b}
Povich et al. 2003\nocite{povich_etal:03}).

The dissociation rate of water into OH and H at 1~AU by solar
radiation (the dominant ionization channel) is $\sim 13\cdot
10^{-6}$~s$^{-1}$ (Budzien et al. 1994\nocite{budzien_etal:94})
and the order of magnitude of a typical ionization/dissociation
rate of cometary molecules and atoms at 1~AU is $10^{-7} -
10^{-5}$~s$^{-1}$ (Huebner et al. 1992\nocite{huebner_etal:92})
which gives a lifetime of the order of days to months. Assuming
the lifetimes decrease as $r^{-2}$, we have lifetimes from 1 to
100~h at $40 R_{\sun}$ and from 0.01 to 1~h at $4 R_{\sun}$.
Comparing the travel time from $40 R_{\sun}$ to $4 R_{\sun}$
($\sim 50$~h) with the apparition rate of new sungrazing comets
(about three comets per week, i.e., one every 50~h) we conclude
that almost permanent although highly fluctuating production of
pickup ions could be observed. Since the final breaking occurs no
closer than $4 R_{\sun}$ and the ionization times at $4 R_{\sun}$
are less than 1~h, we estimate that the seed material can approach
not closer than $2 R_{\sun}$ from the Sun's center. According to
Mann et. al~(2004\nocite{mann_etal:04}) all dust grains must
evaporate before reaching 1.4--2~$R_{\sun}$. Thus, we conclude
that the source region for PUI from sungrazers extends to $\sim 2
R_{\sun}$.

\subsubsection{PUI propagation and detection}

Pickup ions from solar system bodies were observed at surprisingly
large distances of millions of kilometers: Gr{\"u}nwaldt et al.
(1997)\nocite{grunwaldt_etal:97} discovered the Venus PUI tail
from the Earth's orbit, Gloeckler et al.
(2000b)\nocite{gloeckler_etal:00b} detected the PUI tail from
Comet C/1996~B2 Hyakutake, and Gloeckler et al.
(2004\nocite{gloeckler_etal:04a}) observed an oxygen and carbon
PUI signal that shows characteristics resembling the sungrazer
signature discussed by Iseli et al. (2002\nocite{iseli_etal:02}).

To determine the prospective region for detection of sungrazer
PUIs, we make the important assumption that PUIs in the solar wind
propagate radially away from the Sun whatever their injection
velocity and heliocentric distance of their injection point.

Trajectories of singular O$^+$ PUIs in the solar wind
(scatter-free approximation) were calculated by Luhmann
(2003)\nocite{luhmann:03}. Among other cases, she considered a
parent oxygen atom, which at 0.3~AU from the Sun has a velocity of
70~km~s$^{-1}$ and becomes ionized. Such parameters correspond
almost exactly to the trajectories of the sungrazing comets at
this heliocentric distance. Luhmann showed that when the pitch
angle is not scattered, the assumption of radial propagation of
PUIs from their injection point is reasonable only for injection
distances of a few AU from the Sun. Inside 1~AU the propagation is
not radial. When starting at 0.3~AU and 70~km~s$^{-1}$, the PUIs
propagate almost perpendicularly to the Sun -- injection point
line, and at 1~AU this angle is $\sim 45$\degr.

On the other hand, it is highly improbable that there is no
scattering of the pitch angle. Pitch angle scattering would tend
to make the trajectories better aligned with the Sun -- injection
site line. Also, one cannot neglect the influence of local
modifications (draping) of the interplanetary magnetic field by
the comet and their consequences for cometary PUI motion (Luhmann
et al. 1988\nocite{luhmann_etal:88}). Gloeckler et al.
(1986)\nocite{gloeckler_etal:86} observed almost radial
propagation of PUIs from Comet 21P/Giacobini-Zinner from a few
million kilometers in what seemed like a magnetic tail and
Gloeckler et al. (2000b)\nocite{gloeckler_etal:00b} saw a close to
radial propagation of PUIs from Comet C/1996~B2 Hyakutake.
Admittedly, much farther-from-radial propagation is reported by
Gloeckler et al. (2004\nocite{gloeckler_etal:04a}), but in this
case the parent body of the observed signal has not been
unambiguously identified and there is a possibility that the PUIs
had been deflected by a CME. All this makes us adopt the simplest
assumption of radial propagation of PUIs (with a few degrees of
broadening) both from the tails and from the breaking up nuclei of
the sungrazing comets.

To find the range of ecliptic coordinates subtended by the
sungrazing comets and their comae, we computed actual trajectories
on the inbound (pre-perihelion) leg of all known LASCO comets from
the Kreutz, Meyer and Marsden \& Kracht groups between 1~AU and
the breakup region at a few solar radii from the Sun and plotted
the envelope (see Figs. \ref{fig3} and \ref{fig4}). For all
comets, we assumed an eccentricity equal to 1 (parabolic orbit),
which -- given eccentricities $\sim 0.999$ of those few comets for
which it was possible to determine -- is a sufficient
approximation inside 1~AU. Additionally, for the Kreutz group we
selected the most probable orbital elements, taking them as peak
values from the histograms of the orbital parameter distribution.
A track of this orbit is superimposed on the plots in Figs.
\ref{fig3} and \ref{fig4} and is supposed to be the region of the
highest flux of the PUIs and ENAs resulting from the cometary
material. We list the parameters of this orbit along with the mean
orbital elements from our sample in Table~1 and in Table
\ref{tab2} we supplement them with the mean values of orbital
parameters of the Meyer and Marsden \& Kracht groups. We construct
our charts of detection regions of sungrazer PUIs as a radial
projection of the source regions on sky. For the Kreutz group, we
show separately the region corresponding to the volume traversed
by the comets themselves and the envelope, assumed to be
5\degr~wide, due to their neutral-gas clouds.

Earth- and an L1-orbiting spacecraft travel at ecliptic latitude 0
and enter the region of PUI flux from the Kreutz comets each year
on about July 30 (DOY 211) and exit it on November 3 (DOY 307).
For the following $\sim 2$ months they are still inside the region
of PUIs from the secondary groups, exiting it on December 26 (DOY
360). During that time, we expect 48 comets each year from the
Kreutz group and 4 from the secondary groups. The passage through
the sky projection of the orbit of the most probable Kreutz comets
occurs on September 28 (DOY 271). These dates and longitude
intervals are given in Table 3 and shown in Fig.\ref{fig3}.

Ulysses was launched when the Earth was in the Meyer, Marsden \&
Kracht region, which it left on January 1, 1991. Then it coasted
outside the regions prone to detection of PUIs from the sungrazer
population until Nov 25, 1994, when it entered the Meyer region,
which it exited on Dec 13, 1994. Then it entered the Marsden \&
Kracht region on Jan 30, 1995 and the Kreutz region on Feb 10,
1995, which it exited on May 14, 1995. This sequence was repeated
during the second revolution of Ulysses around the Sun, entering
the Meyer region on Feb 02, 2001 and exiting it on Feb 23, 2001,
and entering the Marsden \& Kracht region on Apr 12, 2001 and the
Kreutz region on Apr 23, 2001, where it stayed until Jul 23, 2001.
Ulysses dates and intervals are given in Table 4 and shown in
Fig.\ref{fig4}.

If some of the inner source of pickup ions is indeed sungrazing
comets, then in the time intervals indicated above one should
expect seasonal changes of the inner source observed from 1~AU
both in the absolute flux and in the chemical composition. Ulysses
should see the sungrazers' inner source on the intervals indicated
in Fig.\ref{fig4}, which correspond well with the dates of PUI
observations discussed by Gloeckler \& Geiss (1998,
2001)\nocite{gloeckler_geiss:98a} \nocite{gloeckler_geiss:01} and
Gloeckler et al. (2000a)\nocite{gloeckler_etal:00a}.

The size of the detection area of the Kreutz comets is about
2.39~sr, i.e.\@ $\sim 0.19$ of the full solid angle. Hence the
flux of all PUIs from the Kreutz comets averaged over time and
the detection area should be $1.6\cdot 10^5$~g~s$^{-1}$~sr$^{-1}$.
Geiss et al. (1996\nocite{geiss_etal:96a}) estimate the oxygen
flux from the inner source as $2\cdot 10^6$~g~s$^{-1}$ in the full
solid angle, which gives $1.6\cdot 10^5$~g~s$^{-1}$~sr$^{-1}$ --
only two or three-fold higher given the mass fraction of oxygen in
a cometary nucleus equal to $\sim 0.3$ (Greenberg \& Li
1999\nocite{greenberg_li:99}) -- and Gloeckler \& Geiss
(1998\nocite{gloeckler_geiss:98a}) give $7\cdot 10^6$~g~s$^{-1}$,
i.e.\@ $5.6\cdot 10^5$~g~s$^{-1}$~sr$^{-1}$ -- a higher value but
still comparable to ours given the large uncertainty of all these
estimates.

\subsection{Energetic Neutral Atoms}

One of the channels of PUI production from the sungrazers'
material is charge exchange between solar wind protons and
cometary neutrals. The reactions potentially important in the
context of H ENA production are the following:
\begin{equation}
\label{reaction1}
{\rm H} + {\rm p} \rightarrow {\rm H}^+_{\rm PUI} + {\rm H_{ENA}},
\end{equation}
\begin{equation}
\label{reaction2}
{\rm O + p \rightarrow O^+_{PUI} + H_{ENA}},
\end{equation}
and
\begin{equation}
\label{reaction3}
{\rm C + p \rightarrow C^+_{PUI} + H_{ENA}},
\end{equation}
where H, O, and C on the left-hand side are the cometary neutrals
and p are the solar wind protons. These reactions are practically
the sole sources of Energetic Neutral Atoms of cometary origin
since all other elements in the solar wind are highly ionized and
cannot produce a neutral even if they exchange charge with a
cometary neutral atom. We estimate the number $N_{\rm ENA}$ of H
ENAs produced by these reactions as a percentage of all atoms
released by the breaking up nucleus and subsequently ionized and
picked up by the solar wind as:
\begin{equation}
\label{eq2}
N_{\rm ENA} = \sum_{A \in \{H, O, C\}} \xi_{\rm A} X_{\rm A}
\end{equation}
where $X_{\rm A}$ is the number abundance of atoms A in the cometary nucleus and
$\xi_{\rm A}$ is the percentage of the charge exchange rate $\beta_{\rm
A,cx}$ of element A to the total ionization rate of this element:
\begin{equation}
\label{eq3}
\xi_{\rm A} = \beta_{\rm A,cx}/\left(\beta_{\rm A,EUV}+\beta_{\rm A,cx}+
\beta_{\rm A,el}\right).
\end{equation}

For the EUV rates we took mean values between solar minimum and
maximum given by Ruci{\'n}ski et al.
(1996\nocite{rucinski_etal:96a}), scaled to the appropriate
distance according to $1/r^2$. To compute the electron impact
rates, we used the solar wind density and temperature model by
K{\"o}hnlein (1996\nocite{kohnlein:96}), with the proton density
scaled up by 10\% to account for electrons from the fully-stripped
elements heavier than hydrogen, and we assumed that the electron
distribution function is mono-Maxwellian. For hydrogen, we
calculated the electron impact rate following Ruci{\'n}ski \& Fahr
(1989\nocite{rucinski_fahr:89}) and for oxygen and carbon
following Phaneuf et al. (1987\nocite{phaneuf_etal:87}). To
compute charge exchange, we used the aforementioned model of solar
wind parameters by K{\"o}hnlein (1996\nocite{kohnlein:96}) and
velocity-dependent charge exchange cross sections between protons
and H, O, and C atoms, respectively from Bzowski
(2001\nocite{bzowski:01b}), Stancil et al.
(1999\nocite{stancil_etal:99}) and Stancil et al.
(1998\nocite{stancil_etal:98a}). We also computed the actual
relative velocity between the typical Kreutz nucleus and the solar
wind, which turned out to be weakly varying with the heliocentric
distance and equal to $\sim 430$~km~s$^{-1}$.

>From the three ionization channels, we assembled the ionization
budget of H, O, and C defined by equations (\ref{eq2}) and
(\ref{eq3}) at a few heliocentric distances between 1 and 0.01~AU.
We found that percentage of the charge exchange rate in the net
ionization rate of these ions in the solar neighbourhood depends
only weakly on the heliocentric distance. For hydrogen, it
increases from $\sim 66$\% at 0.1~AU to $\sim 70$\% at 0.01 AU,
for oxygen from $\sim 27$\% to $\sim 30$\% and for carbon from
$\sim 7$\% to $\sim 10$\%. Given the atomic abundances mentioned
earlier, this yields the H ENA injection rate at the level of
$\sim 40$\% of all PUIs injection rate in the nuclei breakup
region.

Since in the charge exchange reactions in question there is no
significant momentum exchange, the newly created H ENAs inherit
the motion of the incident protons and escape. If this happens in
the coma, the geometry is simple and the propagation close to
radial (a ``pencil beam''). Raymond et al.
(1998\nocite{raymond_etal:98}) and Uzzo et al.
(2001\nocite{uzzo_etal:01}) point out a different possibility.
They observed bright sungrazers close to and in the breakup region
in the Lyman-$\alpha$ line using UVCS on SOHO. They observed a
cometary linewidth of $\sim 200$~km~s$^{-1}$, which they interpret
not as caused by the H atoms emitted by the nucleus and travelling
almost parallel to the comet towards the Sun, because those at
these distances from the Sun are already Doppler-shifted outside
the spectral range of the solar line, but as shocked solar wind
protons, which were subsequently neutralized by charge-exchange
with the new H, O, and C atoms from the nucleus. The velocity
direction of these particles has been distributed widely by the
cometary shock and some of them are inside the spectral range of
the solar line. If such a scenario is realized, then a
``shrapnel'' geometry of propagation of these atoms is more
appropriate, with atoms dispersed into the full antisolar
hemisphere. The beam would be much widened at the expense of its
flux magnitude.

Solar wind accelerates in the breakup region from $\sim 50$~km~s$^{-1}$ at
$2.2\,R_{\sun}$ to $\sim 400$~km~s$^{-1}$ (K{\"o}hnlein
1996\nocite{kohnlein:96}), so the H atoms starting off at 50~km~s$^{-1}$ at
$2.2\,R_{\sun}$ do not have enough energy to reach 1~AU without support from the
solar Lyman-$\alpha$ radiation pressure. The Doppler width of this self-reversed
line is about 120 km~s$^{-1}$ (Lemaire et al. 2002\nocite{lemaire_etal:02}) and
the magnitude of the force exerted by solar photons exceeds the solar gravity
force for radial velocities between $\sim 30$~km~s$^{-1}$ and $\sim 75$~km~s$^{-1}$. Hence,
after very rapid acceleration from the start speed to the maximum speed at which
the radiation pressure exceeds solar gravity, these atoms ``surf'' on the
radiation pressure away from the Sun at a constant speed of $\sim
75$~km~s$^{-1}$ ($\sim 30$~eV).

The atoms created farther away from the Sun have a larger initial
speed because the seed solar wind protons are faster, and they are
much less decelerated by solar gravity: those launched at
$4\,R_{\sun}$ have a speed of 120~km~s$^{-1}$ (75 eV) and
immediately feel the solar radiation pressure, and those launched
at 0.1~AU have a speed of 320~km~s$^{-1}$ (0.53~keV) and at 1~AU
are slowed down to $\sim 290$~km~s$^{-1}$ (0.44~keV), feeling no
radiation pressure because they are Doppler-shifted away from the
solar line.

Once created, the cometary H ENAs are subjected to ionization loss
processes from all three hydrogen destruction channels. The EUV
ionization and electron impact rates are similar to those of the
original cometary H atoms. Charge exchange operates differently
because of the different relative speeds between the local solar
wind and the ballistic H ENAs. Since the speed of a newly-created
H ENA is determined by the local solar wind speed and is further
reduced by solar gravity while solar wind accelerates with the
increase of heliocentric distance, the charge exchange rate is
highest in the case of H ENAs created closest to the Sun. We
estimated the survival probability of an H ENA injected at the
typical breakup distance of 0.05~AU and, for comparison, at 0.1~AU
(correspondingly, 10 and 20~$R_{\sun}$). It turns out that
practically all atoms launched at 0.05~AU will be ionized (the
survival probability will be $\sim 4 \cdot 10^{-4}$), while out of
those launched at 0.1~AU $\sim 20$\% will be able to reach Earth's
orbit.

Destruction of H ENAs created close to the Sun has another
interesting consequence. When ionized, they will have a lower
velocity than the surrounding solar wind and consequently they
will be picked up. However, their injection velocity will not be
0, so in the solar wind-related phase space they will be
relatively close to the solar wind core, while retaining their
pickup characteristics. Consequently, they will augment the
original PUI population, created much closer to the Sun, and mimic
PUIs created in the immediate solar neighborhood. Thus ENAs
produced from charge exchange between solar wind protons and
cometary neutrals will be another source of the inner source PUIs.
Effectively, these will be protons extracted from the solar wind
core and transported in phase space to the inner source PUI
region.

The mean rate of H ENA production will be $8 \cdot 10^{26}$ H
atoms per second (40\% of the PUI production rate). Assuming the
``shrapnel'' geometry of initial propagation and considering the
low probability of survival from the breakup region to 1~AU, we
can expect a mean H ENA flux at Earth equal to just $10^{-4}$
at~s$^{-1}$~cm$^{-2}$. In reality, the mean flux is not a good
measure: we expect one comet every two days and the breakup takes
about 2 hours. Therefore, actual instantaneous fluxes should be
$\sim 25$-fold higher, i.e. $2.5 \cdot 10^{-3}$
at~s$^{-1}$~cm$^{-2}$, however the durations should be just about
2 hours. For comparison, the H ENA flux from interstellar hydrogen
at the offset angle $\sim 120\degr$ from the upwind direction is
expected to be of the order of 1 to 50 at s$^{-1}$~cm$^{-2}$,
depending on the phase of solar cycle (Bzowski et al.
1996\nocite{bzowski_etal:96}). Hence, the H ENAs from comets would
be observable at Earth only close to solar maximum, and with luck.
However, they will be a source of the extra 25\% of the sungrazer
PUI flux on top of the PUI originating from the cometary material
itself.

\section{Discussion}

The suggestion that comets might be responsible for the inner
source of pickup ions was proposed quite early (Geiss et al.
1995\nocite{geiss_etal:95a}\nocite{geiss_etal:96a}), although the
primary mechanism envisaged was loss of volatile components at
larger distances from the Sun. Frequently appearing small comets
from isotropic directions would be hardly distinguishable from an
extended source of particles (Gloeckler \& Geiss
1998\nocite{gloeckler_geiss:98a}).

The composition of the inner source shows depletion of hydrogen
with respect to solar wind abundances. Schwadron \& Geiss
(2000)\nocite{schwadron_geiss:00} extensively discuss possible
mechanisms of this depletion under the assumption that most of the
inner source hydrogen is released from dust grains as H$_2$
molecules. If, however, we take the sungrazing comets as the inner
source, the problem of hydrogen underabundance is strongly
alleviated, even with the extra PUIs from the reionized cometary H
ENAs taken into account.

The chemical composition of the inner source PUI inferred by
Gloeckler \& Geiss (1998)\nocite{gloeckler_geiss:98a} and
Gloeckler et al. (2000a)\nocite{gloeckler_etal:00a} and discussed
extensively by Schwadron et al. (2000)\nocite{schwadron_etal:00}
shows, among others, helium and neon. Noble gases are not expected
in ``old'' dust grains near the Sun and hence a mechanism of
absorption of solar wind neon ions to the dust grains and its
further release after recombination was proposed. Sungrazers do
not alleviate this problem directly: very little, if any at all of
atoms of noble gases were found among cometary volatiles (Geiss et
al. 1999, \nocite{geiss_etal:99} Bocl{\'e}e-Morvan et al.
2004\nocite{boclee-morvan_etal:04}). Since, however, they supply
dust to the immediate solar neighbourhood then the same mechanism
that neutralizes solar wind neon on solar system dust grains
should operate on the dust grains from comets.

The evolution of PUIs from the inner source both with respect to
the phase space and to the charge state was discussed by Schwadron
et al. (2000)\nocite{schwadron_etal:00}. They point out that
during the drift away from the Sun the PUI should experience more
or less adiabatic cooling which would make it increasingly
difficult to observe them with the increasing distance to the Sun.
Therefore, in the case of Ulysses, the best time seems to be when
the spacecraft is close to its perihelion, i.e., in the portion of
the orbit close to the ``fast latitude scan''.

In the estimates of the PUI flux we took an average production
rate. In reality, the sungrazer PUI flux will be highly
fluctuating both in time and in space. On the one hand, one can
expect that even when a comet is as close to the Sun as 0.2~AU,
the source region for PUIs should be large enough to result in a
few days' duration of the PUI signal at the Earth or Ulysses orbit
(Gloeckler et al. 2004\nocite{gloeckler_etal:04a}). Indeed,
Gloeckler et al. (2000b\nocite{gloeckler_etal:00b}) observed with
Ulysses the PUI tail of Comet C/1996~B2 Hyakutake when it was at
$\sim 0.35$ AU from the Sun and the signal (apart from a $\sim 1$
day peak) seems to last for 5 -- 6 days. On the other hand, the
average apparition rate for the Kreutz group is one every other
day (and of the secondary groups one per month) and the apparition
times show a Poisson distribution. Therefore it cannot be ruled
out that the inner source of PUIs from the Kreutz sungrazers will
be intermittently drying out and that the PUIs from the secondary
groups will not be detected at all.

Further, we assumed a radial propagation of the populations
created from the cometary material, which need not be true. While
PUIs from large sungrazers can probably be observed as
well-defined signals (Iseli et al. 2002\nocite{iseli_etal:02}),
the fluxes from the much more frequent smaller objects can be so
low that they will be registered as singular events, hard to
notice on a case by case basis and showing up only when a
comparison of long-time averages is made. A good test of the
reality of the cometary character of the inner source would be to
average available PUI and possibly ENA observations over the time
intervals proposed in this paper and to compare the results with
an averaging performed outside these time frames.

\section{Conclusions}

We suggest that a considerable portion of the inner source of
pickup ions may originate from the SOHO sungrazing comets,
especially from the Kreutz group. We show that the actual rate of
inflow of these comets may be as high as one per two days. The
corresponding total mass production rate would be $3.1 \cdot
10^4$~g~s$^{-1}$, released into the solar wind between $\sim
40$~R$_{\sun}$ and $\sim 2\,R_{\sun}$ and emitted into a
restricted area of $\sim 0.19$ of the full volume angle, so that
the flux is $1.6 \cdot 10^5$~g~s$^{-1}$~sr$^{-1}$. Together with
the re-ionized H ENAs created by charge exchange between cometary
atoms and solar wind protons, this gives a mean PUI particle flux
equal to $1.2\cdot 10^{27}$~s$^{-1}$~sr$^{-1}$, which at 1~AU is
equivalent to 12.5~s$^{-1}$~cm$^{-2}$. The intensity of
sungrazer-related populations should be strongly variable both in
time and in space. Because of the geometry of the orbits, the
pickup ions should be observable by an Earth- or L1-orbiting
spacecraft between end of July and the end of the year. Ulysses
should have been within the detection area since its launch until
the end of 1990 (but because of the scarcity of this source it
could have missed them), and then during its two fast latitude
scans between $\sim -25$\degr\, and $\sim 40$\degr\, ecliptic
latitude, with a plunge into a low detection-probability area
between $\sim -75$\degr\, and $\sim -60$\degr\, latitude.
Measurements of the inner source pickups carried out within these
areas should show different intensities and characteristics than
when performed elsewhere. To our knowledge, the cometary
hypothesis is the only one that predicts a distinct ecliptic
longitude effect and as such should be easily verifiable by
inspection of existing PUI observations from Ulysses and 1~AU
spacecraft, as Wind and ACE.

\begin{acknowledgements}
The authors gratefully acknowledge discussions with Grzegorz Sitarski, Andrzej
Czechowski, and Stan Grzedzielski from the Space Research Centre PAS, and
Eberhard M{\"o}bius from the University of New Hampshire. This research was
supported by the Polish State Committee for Scientific Research Grant
1~P03D~009~27.
\end{acknowledgements}

\bibliographystyle{aa}
\bibliography{iplbib}
\end{document}